\documentclass[a4paper,11pt]{article}
\usepackage{jheppub} 
\usepackage[section]{placeins}
\usepackage[T1]{fontenc} 
\usepackage{graphicx}
\usepackage{amsmath}
\usepackage{bm}
\usepackage{subfigure}
\usepackage{amssymb}
\usepackage{mathrsfs}
\usepackage{braket}
\usepackage{multirow,booktabs}
\usepackage{comment}

\newcommand{\comments}[1]{}

\usepackage{slashed}
\usepackage{epsfig,psfrag,rotating,soul}
\usepackage{rotfloat}
\usepackage{empheq}  

\allowdisplaybreaks

\title{\boldmath  Non-Classicality in $tW$ system via negativity and geometric quantum discord}

\author[a,b]{Amir Subba }

\author[a,b]{, Yu Shi }

\affiliation[a]{ Wilczek Quantum Center, Shanghai Institute for Advanced Studies, Shanghai 201315, China }  

\affiliation[b]{
University of Science and Technology of China, Hefei 230026, China }

\emailAdd{amirsubba@ustc.edu.cn}

\emailAdd{yu\_shi@ustc.edu.cn}  

\abstract{
We perform a phenomenological study of non-classical correlations in the mixed-dimensional qubit-qutrit system realized by leading order $pp\to tW$ production at $\sqrt{s}=13$ TeV in the leptonic final state. The two missing neutrinos are reconstructed with conditional normalizing flows trained on simulated detector-level objects, and the joint spin density matrix $\rho_{tW}\in\mathcal{H}_2\otimes\mathcal{H}_3$ is obtained in a generalized Bloch representation from the angular distributions of the leptonic decay products. Non-classical correlations are quantified using the Peres--Horodecki negativity and geometric quantum discord. We quantify the correlations at different region of $m_{tW}$ at the reconstruction level within the Standard Model and the impact of two reducible backgrounds viz. $t\bar{t}$ and $WWj$ processes, tagged using boosted decision trees. We further study the behavior of correlations in presence of anomalous chromomagnetic and chromoelectric dipole moments of the top quark. The sensitivity reach of the quantum observables along with polarization and spin correlation parameters are reported for current LHC and future HL-LHC scenario. 
}

\begin{document}

\maketitle

\section{Introduction}
\label{sec:intro}

The conceptual foundation of quantum correlations dates back to the Einstein–Podolsky–Rosen (EPR) paradox~\cite{Einstein:1935rr}, which questioned the completeness of quantum mechanics by highlighting non-classical correlations between spatially separated systems.  Schr\"odinger identified quantum entanglement (QE) as the defining feature of such correlations, emphasizing the non-separability of composite quantum states~\cite{Schrdinger1935}. QE implies that the joint state of a composite system cannot be expressed as a convex combination of product states and therefore lacks a classical probabilistic interpretation.

The operational distinction between quantum and classical correlations was formalized by Bell~\cite{Bell:1964kc}, who derived inequalities satisfied by any local hidden-variable theory. Violation of Bell inequalities provides a device-independent signature of non-local quantum correlations. Experimental tests, beginning with the pioneering measurements of Aspect~\emph{et al.}~\cite{Freedman:1972zza,Aspect:1982fx} and culminating in loophole-free demonstrations~\cite{Hensen:2015ccp,Giustina:2015yza,Stevens:2015awv}, established the incompatibility of quantum mechanics with local realism. Beyond foundational significance, QE is now recognized as a key resource in quantum information science, enabling quantum communication~\cite{Ursin:2006jya}, quantum cryptography~\cite{Yin:2020rtd}, quantum computation~\cite{Raussendorf:2001snk,Nielsen:2012yss}, and precision metrology~\cite{Nielsen:2012yss,Giovannetti:2004cas,Toth:2014msl}.  

Despite its central role, QE does not exhaust all forms of non-classical correlations. A broader characterization is provided by quantum discord~\cite{Ollivier:2001fdq,Henderson:2001wrr}, which arises from the inequivalence between two quantum generalizations of classical mutual information. Quantum discord quantifies the minimum disturbance induced by local measurements and can be non-zero even for separable states. This reveals that quantum correlations extend beyond non-separability and may persist in mixed states where QE measures vanish. Discord has been identified as a resource in deterministic quantum computation with one qubit ~\cite{Datta:2007err,Datta:2009jfh}, quantum state discrimination~\cite{Modi:2010ubn,Piani:2011bjd}, quantum metrology~\cite{Girolami:2013uta}, and noise-assisted quantum protocols~\cite{Lanyon:2008ker}.

QE in high energy physics has played an important historic role~\cite{shiwu,shiroad,shihistoric}. Moreover, 
QE in pseudoscalar mesons ~\cite{shiteleportation,shiteleportation2,huangshi,shicp,shicp2,huangshi2,shiyang,shiyangleggett,shiyangbell} and in relativistic quantum field theory~\cite{shifield} has been studied for many years. 

Quantum correlations at colliders have been studied predominantly through entanglement, with early and subsequent studies focusing on bipartite spin systems. A collider experiment cannot select measurement settings and measure a particle's spin directly, since the decay time is far too short, e.g., $\sim10^{-25}$~s for both the top quark and the $W$ boson. The spin density matrix is instead reconstructed from the angular distributions of the decay products, with the Standard Model~(SM) providing the matching to the underlying spin observables. This method has been employed at $\sqrt{s}=13$ TeV LHC to discover the quantum entanglement in $t\bar{t}$ pairs~\cite{CMS:2024pts,ATLAS:2023fsd} and recently ATLAS~\cite{y1nh-1b82} reported evidence of QE on $H \to ZZ^\star \to 4\,\ell$ final state at $4.7$ standard deviations.

For two spin-$1/2$ particles, i.e. two-qubit systems, QE has been investigated extensively in $t\bar t$ production, including studies of its observability at the LHC, Bell-inequality violation, post-decay effects, and its sensitivity to physics beyond the Standard Model~\cite{Afik:2020onf,Cheng:2024btk,Dong:2023xiw,Han:2023fci,Aguilar-Saavedra:2023hss,Fabbrichesi:2021npl,Severi:2021cnj,Afik:2022kwm,Aoude:2022imd,Aguilar-Saavedra:2022uye,Fabbrichesi:2022ovb,Dong:2023xiw,Cheng:2023qmz,Maltoni:2024tul,Aguilar-Saavedra:2024hwd,Aguilar-Saavedra:2024vpd,Lamba:2026sni,A:2026pug,CamachoJuarez:2026sss,Jiang:2026sfw}. Entanglement in other fermionic two-qubit systems has also been explored, notably in $\tau^+\tau^-$ production and in Higgs-boson decays to $\tau^+\tau^-$~\cite{Fabbrichesi:2022ovb,Ashby-Pickering:2022umy,Altakach:2022ywa,Ma:2023yvd,Ehataht:2023zzt}, $\Lambda\bar{\Lambda}$ pair~\cite{Wu:2024mtj} and $b\bar{b}$ final state~\cite{Afik:2025grr}. Beyond two-qubit systems, entanglement has been studied for pairs of spin-1 particles, corresponding to bipartite $1\otimes1$ or two-qutrit systems, in diboson production and vector-boson scattering~\cite{Aoude:2023hxv,Morales:2023gow,Bi:2023uop,Subba:2024mnl,De:2025dpo,Goncalves:2026njf,Aguilar-Saavedra:2026wuq,Aguilar-Saavedra:2022wam,Aguilar-Saavedra:2022mpg,Bernal:2024xhm,Ashby-Pickering:2022umy,Severi:2022qjy,Fabbrichesi:2023cev,Bernal:2023ruk,Bi:2023uop}. Mixed-dimensional bipartite systems, such as the $tW$, $\Lambda\phi/K^{\star\,0}$ final state, provide examples of qubit--qutrit systems and have also been investigated~\cite{Subba:2024mnl,Kanachova:2026ywy}. More recently, multipartite entanglement has been considered in processes such as $t\bar t Z$ production and Higgs decay~\cite{Subba:2024mnl,Cao:2026mza,Goncalves:2026nnx,Banacki:2026msu,Aguilar-Saavedra:2024whi}, extending collider studies beyond the bipartite setting.

Beyond QE, other measures of non-classical correlations have also
been explored at colliders, including quantum
steering~\cite{Afik:2022dgh,Altakach:2022ywa,Aguilar-Saavedra:2026grl,Zhang:2026wvn,Wu:2025dds}
and magic~\cite{White:2024nuc,Li:2026zum,Chu:2026yxm}. Quantum discord
offers a more general characterization of non-classicality, as it can
remain nonzero even for separable states; its collider phenomenology
has been studied in $t\bar{t}$
production~\cite{Afik:2022dgh,Han:2024ugl,Subba:2026nzs,Cheng:2026ktp,Zhang:2026wvn,Wu:2025dds}. 

A direct evaluation of quantum discord (QD) based on its entropic definition requires an optimization over local measurements and is computationally demanding for higher-dimensional systems. The closed-form analytic expressions for discord are available for maximally marginal density matrices~\cite{Luo:2008kvk} and \(X\)-states~\cite{Chen:2011spg}. The closed form for QD for the \(t\bar{t}\) state within the SM has been presented in Ref.~\cite{Han:2024ugl}. The alternate form of QD can be obtained via geometric quantum discord (GQD), defined as the minimal Hilbert--Schmidt distance between a given state and the set of classical--quantum states~\cite{Dakic:2010xfz}.

While quantum discord in collider studies has so far been explored predominantly for $1/2\otimes1/2$ systems, most notably through $t\bar{t}$ pair production, its extension to mixed-dimensional $1/2\otimes1$ qubit--qutrit systems remains largely unexplored. The associated production of a single top quark with a $W$ boson provides a natural realization of such a bipartite quantum system: the top-quark spin defines a qubit, while the $W$-boson polarization spans a three-dimensional Hilbert space. Motivated by this, we extend the study of quantum discord, via its geometric formulation, to the $tW$ system.

The $tW$ process has been measured at $\sqrt{s}=13$ TeV by CMS~\cite{CMS:2022ytw} with an inclusive cross section $\sigma =  79.2 \pm 0.9\,(\mathrm{stat})^{+7.7}_{-8.0}\,(\mathrm{syst})\pm 1.2\,(\mathrm{lumi})$ pb, and by ATLAS~\cite{ATLAS:2024ppp} with $\sigma = 75^{+15}_{-14}$ pb in the leptonic final state, establishing it as a well-characterized process at the LHC. In this work, we analyze the non-classicality in the $tW$ system produced at the 13 TeV LHC at leading order. We consider the leptonic decay of the top quark and the $W$ boson, resulting in a final state with two charged leptons, one $b$-jet, and two missing neutrinos. We employ the normalizing-flows architecture described in Ref.~\cite{Raine:2023fko} to reconstruct the missing neutrinos. Owing to the short lifetime of the top quark and the chiral structure of weak interactions, the leptonic decay products act as efficient spin analyzers, enabling reconstruction of the production density matrix from angular distributions~\cite{Boudjema:2009fz,Rahaman:2023pte}. The $tW$ density matrix is constructed from the joint angular distribution of the final-state charged leptons in the rest frames of the top quark and the $W$ boson. To quantify non-classical correlations, we employ the Peres--Horodecki negativity of the partially transposed density matrix and the GQD, and examine their kinematic dependence on the invariant mass $m_{tW}$. The impact of reducible backgrounds is studied in the presence of $t\bar{t}$ and $WWj$ processes, with event selection achieved using boosted decision trees. We further study the modification of these correlations in the presence of anomalous chromomagnetic and chromoelectric dipole moments of the top quark. We highlight the sensitivity of the quantum observables to the anomalous dipole moments at different benchmark values of integrated luminosities and systematic uncertainties at the LHC.

The remainder of this paper is organized as follows. In sec.~\ref{sec:formalism}, we discuss the formalism of reconstructing the qubit--qutrit density matrix for the $tW$ system from the angular distribution of decayed charged leptons. Sec.~\ref{sec:reco} discuss the collider simulation and reconstruction of two missing neutrinos using normalizing flows. Section~\ref{sec:GQD_tW} introduces quantum observables viz. Peres-Horodecki negativity and geometric quantum discord and presents the Standard Model predictions for disjoint region of invariant mass of top-$W$ system at the level of parton and reconstruction. In sec.~\ref{sec:background}, we discuss the impact of inclusion of reducible background on the estimation of quantum observables. In sec.~\ref{sec:anom}, we investigate the dependence of negativity and geometric discord on anomalous top-quark dipole moments along with discussion on sensitivity reach at current and projected HL-LHC luminosity and two benchmark values of systematic uncertainty.  And finally, we conclude in sec.~\ref{sec:conclusion}.

\section{\texorpdfstring{The $tW$ density matrix}{Reconstruction of tW state}}
\label{sec:formalism}

We consider associated single-top production in proton--proton collisions,
\begin{align}
  p p \;\to\; t(p_t, \lambda_t) + W(p_W, \lambda_W),
\end{align}
where $\lambda_t \in \pm 1/2$ and $\lambda_W \in 0,\pm1$ denotes the helicity state of the top quark and $W$ boson. At leading order, the dominant partonic sub-processes are
\(bg \to tW^-\) and its charge conjugate process.

The spin and polarization information of the produced $tW$ system is encoded in the production density matrix,
\begin{align}
    \rho_{\lambda_t \lambda_W ; \lambda_t^\prime \lambda_W^\prime}
    \propto \sum \mathcal{M}_{\lambda_t \lambda_W}\mathcal{M}^{\ast}_{\lambda_t^\prime \lambda_W^\prime},
\end{align}
where $\mathcal{M}_{\lambda_t \lambda_W}$ denotes the helicity amplitude for the partonic sub-process, and the sum runs over initial-state degrees (spin, color) of freedom. Averaging over the unpolarized initial partons and integrating over unresolved kinematic variables renders the reduced density matrix of the $tW$ system generically mixed.

The top quark decays almost exclusively via
\(t \to b W^+ \to b \ell^+ \nu_\ell,\) before hadronization, such that its spin information is fully preserved in the angular distributions of its decay products. Similarly, the polarization of the associated $W$ boson can be reconstructed from its leptonic decay due to its chiral coupling. The combined leptonic decay chain therefore acts as an efficient analyzer of the joint spin--polarization state of the $tW$ system.

From the perspective of quantum information theory, the top quark spin defines a two-level system (qubit), while the $W$ boson polarization spans a three-dimensional Hilbert space (qutrit). The complete quantum state of the $tW$ system is thus described by a qubit--qutrit density matrix, $\rho_{tW} \in \mathcal{H}_2 \otimes \mathcal{H}_3$.

In the Bloch-like representation for qubit--qutrit systems, the density matrix can be written as~\cite{Rahaman:2021fcz,Subba:2024mnl}
\begin{align}
    \label{eq:rho_tw}
    \rho_{tW} &= \frac{1}{6}\,\Big[
    \mathbb{I}_2 \otimes \mathbb{I}_3
    + \vec{P}^t \cdot \vec{\sigma} \otimes \mathbb{I}_3
    + \frac{3}{2}\,\mathbb{I}_2\otimes \vec{P}^W\cdot\vec{S} +\sqrt{\frac{3}{2}}\mathbb{I}_2 \otimes T^W_{ij}\,\left(S_iS_j+S_jS_i\right) \nonumber\\&+\frac{3}{2}PP^{tW}_{ij}\sigma_i \otimes S_j + \sqrt{\frac{3}{2}}PT^{tW}_{ijk}\sigma_i \otimes \left(S_jS_k+S_kS_j\right)
    \Big],
\end{align}
where $\vec{\sigma}$ are the Pauli matrices acting on the top-quark spin space, $\vec{S}$ are the spin-1 operators generating $\mathrm{SU}(3)$ in the $W$-boson polarization space, $\vec{P}^t$ and $\{\vec{P}^W,T^W\}$ denote the local Bloch vectors of the top quark and the $W$ boson, respectively. The spin correlation between the top and $W$ boson are group in terms of nine vector-vector ($PP^{tW}$) and $15$ vector-tensorial ($PT^{tW}$) correlations.

Each component of $\vec{P}^{t/W}$, $T^W, PP^{tW}$, and $PT_{tW}$ can be expressed as a linear combination of the underlying density matrix elements by performing the trace with appropriate operators. For e.g., the three components of $\vec{P}^t$ can be derived as
\begin{align}
    P^t_i  = \mathrm{Tr}\left[\rho_{tW}.(\sigma_i\otimes \mathbb{I}_3)\right].
\end{align}
Moreover, these parameters can be reconstructed experimentally from the joint angular distributions of the leptonic decay products, which is given as~\cite{Boudjema:2009fz}
\begin{align}
    \label{eq:diff}
    \frac{1}{\sigma}\frac{d^2\sigma}{d\Omega_{l^+}d\Omega_{l^-}} &= \frac{3}{8\pi^2} \sum_{\lambda_t,\lambda_t^\prime,\lambda_W,\lambda_W^\prime}\rho_{tW}\Big(\lambda_t,\lambda_t^\prime,\lambda_W,\lambda_W^\prime\Big)\times \Big[\Gamma_t(\lambda_t,\lambda_t^\prime) \otimes \Gamma_W(\lambda_W,\lambda_W^\prime)\Big].
\end{align}
Here, $\Gamma_{t/W}$ are the normalized decay density which are the function of polar and azimuth orientation of decayed fermions in the rest frame of the top quark and $W$ boson. The parameters of the density matrix are then obtained by partially integrating the above Eq.~\eqref{eq:diff} w.r.t to $\theta$ and $\phi$ of final decayed fermions.

The three vector polarizations of the top quark are obtained via asymmetries in the angular functions of decayed fermion ($f_t$) as~\cite{Rahaman:2023pte}
\begin{align}
    \label{eq:top_pol}
    A[P^t_{(1)}] &\equiv \left(\int_{\theta_{f_t}=0}^\pi \int_{\phi_{f_t}}^{\pi/2}-\int_{\theta_{f_t}=0}^\pi \int_{\phi_{f_t}}^{3\pi/2}\right)d\Omega_{f_t}\,\Big(\frac{1}{\sigma}\frac{d\sigma}{d\Omega_{f_t}}\Big)= \frac{1}{2}\alpha_t\,P^t_{(1)}\\
    A[P^t_{(2)}] &\equiv \left(\int_{\theta_{f_t}=0}^{\pi/2} \int_{\phi_{f_t}}^{\pi}-\int_{\theta_{f_t}=0}^\pi \int_{\phi_{f_t}=\pi}^{2\pi}\right)d\Omega_{f_t}\,\Big(\frac{1}{\sigma}\frac{d\sigma}{d\Omega_{f_t}}\Big)= \frac{1}{2}\alpha_t\,P^t_{(2)}\\
    A[P^t_{(3)}] &\equiv \left(\int_{\theta_{f_t}=0}^{\pi/2} \int_{\phi_{f_t}}^{2\pi}-\int_{\theta_{f_t}\pi/2}^\pi \int_{\phi_{f_t}=0}^{2\pi}\right)d\Omega_{f_t}\,\Big(\frac{1}{\sigma}\frac{d\sigma}{d\Omega_{f_t}}\Big)= \frac{1}{2}\alpha_t\,P^t_{(3)}
\end{align}
Here, $\alpha_t$ is the spin analyzing power of decayed fermion which quantifies the ability of fermion to faithfully reconstruct the polarization state of the mother top quark. For the decay $t \to bW^+(l^+\nu_l)$ through the vertex structure $\bar{f}\gamma^\mu (C_LP_L+C_RP_R)f^\prime V_\mu$ ($P_{L/R} \equiv 1/2 (1\mp \gamma_5)$) and $C_i$ being the chiral coupling, the $\alpha_t$ in the massless limit of final fermions is given by~\cite{Boudjema:2009fz}
\begin{align}
    \alpha_t = \frac{(C_R^2-C_L^2)}{(C_R^2+C_L^2)} = -1.
\end{align}
The polarization of the $W$ boson are computed from the asymmetries associated with the decayed fermion ($f_W$), for e.g, the tensorial polarization $T^{W}_{(12)}$ is given as
\begin{align}
    A[T^W_{(12)}] &\equiv \left(\int_{\theta_{f_W}=0}^\pi\int_{\phi_{f_W}}^{\pi/2}-\int_{\theta_{f_W}=0}^\pi\int_{\phi_{f_W}=\pi/2}^{\pi}+
    \int_{\theta_{f_W}=0}^\pi\int_{\phi_{\phi_W}=\pi}^{3\pi/2}-\int_{\theta_{f_W}=0}^\pi\int_{\phi_{f_W}={3\pi/2}}^{2\pi}
    \right)\nonumber\\&\times d\Omega_{f_W}\Big(\frac{1}{\sigma}\frac{d\sigma}{d\Omega_{f_W}}\Big),\nonumber\\
    &= \frac{2}{\pi}\sqrt{\frac{2}{3}}(1-3\delta_W)T^W_{(12)}.
\end{align}
The spin analyzing power $\alpha_W$ becomes -1 and $\delta_W \to 0$ in the high energy limit~\cite{Boudjema:2009fz}. Similarly, the spin-correlation parameters $PP_{ij}^{tW}$ and $PT_{ijk}^{tW}$ in Eq.~\eqref{eq:rho_tw} are obtained from the joint angular distribution of decayed fermions of both top quark and $W$ boson. Thus, the polarization and spin correlation parameters are related to asymmetries upto a proportionality factor where the asymmetries can be computed numerically as
\begin{align}
    A_{ij}^{tW} = \frac{\sigma(c_i^{f_t}\,c_j^{f_W} > 0) - \sigma(c_i^{f_t}\,c_j^{f_W} < 0)}{\sigma(c_i^{f_t}\,c_j^{f_W} > 0) + \sigma(c_i^{f_t}\,c_j^{f_W} < 0)},\, i\,j \in \{1,\cdots,8\}.
\end{align}
The angular functions associated with correlators $c_i$ are listed in Table~\ref{tab:corrasymm}. For detailed discussion and methodology, we refer the reader to Refs.~\cite{Rahaman:2023pte,Boudjema:2009fz}.
\begin{table}[!h]
    \centering
    \renewcommand{\arraystretch}{1.2}
    \caption{List of asymmetries and the associated angular functions that are used to numerically compute the polarization and spin correlation in $tW$ system. The angular functions for three vector polarization $(A_i,i\in\{1,2,3\})$ for top quark and $W$ boson are same.}
    \label{tab:corrasymm}
    \begin{tabular*}{0.47\textwidth}{@{\extracolsep{\fill}}ccc@{}}\hline
     Asymmetries  & Correlators & Functions  \\
    \hline
      $A_1$   & $c_1$ & $\sin\theta\cos\phi$\\
      $A_2$   & $c_2$ & $\sin\theta\sin\phi$\\
      $A_3$   & $c_3$ & $\cos\theta$\\
      $A_4$   & $c_4$ & $\sin^2\theta\sin(2\,\phi)$\\
      $A_5$   & $c_5$ & $\cos\theta\cos\phi$\\
      $A_6$   & $c_6$ & $\cos\theta\sin\phi$\\
      $A_7$   & $c_7$ & $\sin^2\theta\cos(2\,\phi)$\\
      $A_8$   & $c_8$ & $\sin(3\,\theta)$\\
      \hline
    \end{tabular*}
\end{table}

In the di-leptonic decay channel of the $tW$ system, 
reconstructing the full production density matrix at hadron colliders is challenging due to the presence of two undetected neutrinos and the convolution with parton distribution functions (PDFs), which obscures the initial-state kinematics. Direct event-by-event reconstruction of the spin density matrix from measured observables is therefore non-trivial.

We also note that a fully realistic phenomenological analysis necessarily requires detector-level event reconstruction. In dileptonic final states, this entails solving an intrinsically under-constrained inverse problem associated with the inference of two missing neutrino momenta from the measured visible kinematics and missing transverse momenta. Such reconstruction effects can, in principle, impact the extracted spin-density matrix and the derived quantum correlation observables. 

Different technique exist in the literature in order to reconstruct multiple missing neutrino in an event. Algorithmic strategies include probabilistic neutrino reweighting~\cite{D0:1997pjc}, algebraic solutions imposing $W$-boson on-shell constraints~\cite{Sonnenschein:2005ed}, ellipse methods based on $m_W$ geometry~\cite{Betchart:2013nba}, $M_{T2}$-bounded inference for pair-produced topologies~\cite{Rahaman:2021fcz}, and conditional density estimation via normalizing flows~\cite{Leigh:2022lpn,Raine:2023fko}. In the current work, we employ regression based on conditional flows network implemented in $\nu^2$-Flows developed in Ref.~\cite{Raine:2023fko}, and is discussed in the next section.

\section{Reconstruction of Neutrinos momenta with $\nu^2$-Flows}
\label{sec:reco}
The reconstruction of neutrino momenta in the
$pp \to tW^- \to b\ell^+\nu_\ell\,\ell^- \bar{\nu}_\ell$ (and its charge conjugated) process poses a fundamental challenge at hadron colliders, as the two neutrinos escape the detector without leaving any direct signal. While the total missing transverse momentum $\slashed{\vec{p}}_T = \vec{p}_T^{\,\nu} + \vec{p}_T^{\,\bar{\nu}}$ is inferred from momentum conservation in the transverse plane, the individual neutrino four-momenta remain under-constrained: the system of two massless neutrinos introduces six unknown momentum components, and the available on-shell and kinematic constraints viz. $$m_W^2 = (p_{\ell^-} + p_{\bar{\nu}})^2,\quad m_t^2 = (p_b + p_{\ell^+} + p_\nu)^2,$$ together with the measured $\slashed{\vec{p}}_T$ provides only four equations to solve. Complete reconstruction of the $tW$ kinematics for the angular analyses underlying quantum tomography of the $tW$ system, therefore demands an approach capable of resolving these ambiguities in a physically consistent manner.


To this end, we employ $\nu^2$-Flows~\cite{Raine:2023fko}, a generative machine-learning framework designed for neutrino momentum reconstruction. Rather than regressing to a single point estimate, $\nu^2$-Flows models the conditional probability distribution $p(\vec{p}_{\nu},\vec{p}_{\bar{\nu}}\mid\mathcal{O})$
over the six neutrino momentum components given the observed final-state objects $\mathcal{O}$. This is achieved through a conditional normalizing flow~\cite{Tabak:2010vsy,Rezende:2015ocs}, which learns an invertible mapping between a simple base distribution and the target neutrino-momentum distribution, conditioned on a learned representation of the event. The conditioning context is extracted by the transformer-based encoder of Ref.~\cite{Raine:2023fko}, which processes the reconstructed event objects, including the charged leptons, jets, and missing transverse momentum. We use the network architecture and implementation provided by the authors in the public $\nu^2$-Flows repository\footnote{\url{https://github.com/rodem-hep/nu2flows}}, with the input observables adapted to our event selection as summarized in Table~\ref{tab:inputs}. The full architecture and training procedure are described in Ref.~\cite{Raine:2023fko}; here we summarize only the aspects relevant to our application.

\begin{table}[ht]
    \caption{The different input observables used as inputs to the feature extraction network.}
    \label{tab:inputs}
    \centering
    \begin{tabular}{r c l}
    \toprule
    Category & Variables & Description\\
    \midrule
    $\slashed{p}_T$ & $\slashed{p}_x$, $\slashed{p}_y$ & Missing transverse momentum 2-vector \\ [2ex]
    \multirow{3}{*}{Leptons} & $p_x^{\ell}$, $p_y^{\ell}$, $p_z^\ell$, $\log E^\ell$ & Lepton momentum 4-vector \\
    & $q^\ell$ & Lepton charge\\
    & $\ell^{flav}$ & Whether lepton is an electron or muon \\ [2ex]
    \multirow{2}{*}{Jets} & $p_x^{j}$, $p_y^j$, $p_z^j$, $\log E^j$ & Jet momentum 4-vector \\
    & $isB$ & Whether jet passes $b$-tagging criteria \\ [2ex]
    Misc & $N_{\text{jets}}$, $N_{\text{bjets}}$ & Jet and $b$-jet multiplicities in the event\\
    \bottomrule
    \end{tabular}
\end{table}

\begin{figure*}[!htb]
  \centering
  \includegraphics[width=0.99\textwidth]{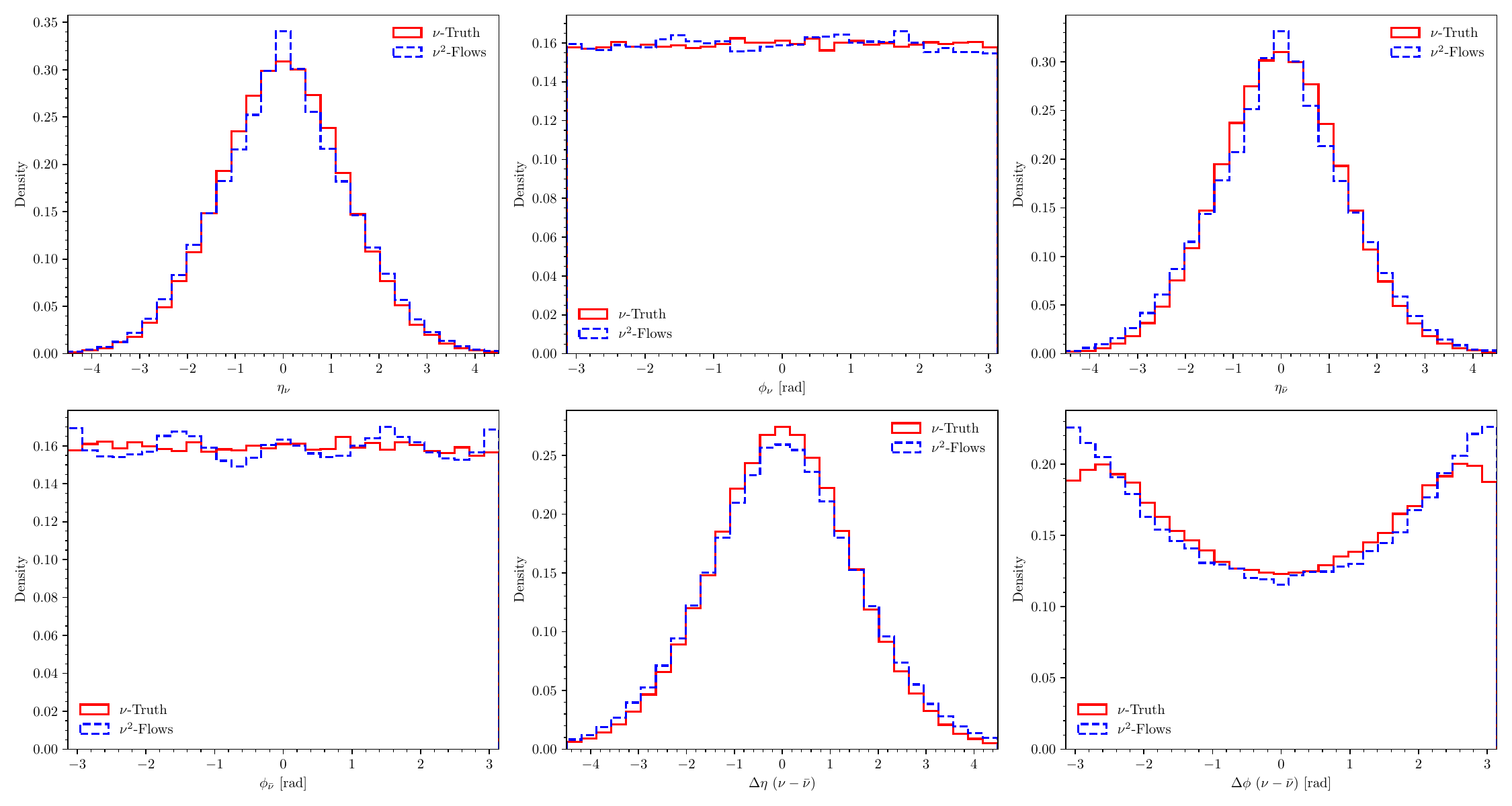}
  \caption{
    Angular distributions of the reconstructed and parton-truth neutrinos.
    Shown are the pseudorapidity $\eta$ and azimuthal angle $\phi$ of the
    neutrino  and antineutrino, together with the angular separations $\Delta\eta$ and $\Delta\phi$ between the neutrino--antineutrino pair.  }
  \label{fig:nu_angular}
\end{figure*}

\begin{figure*}[!htb]
  \centering
  \includegraphics[width=0.495\textwidth, height=6.5cm]{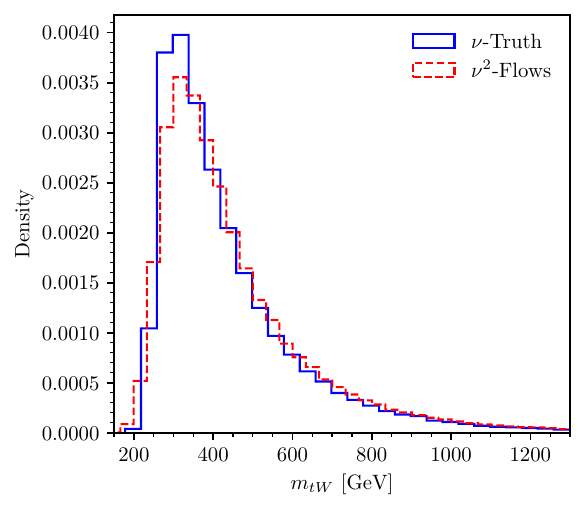}
  \includegraphics[width=0.495\textwidth, height=6.5cm]{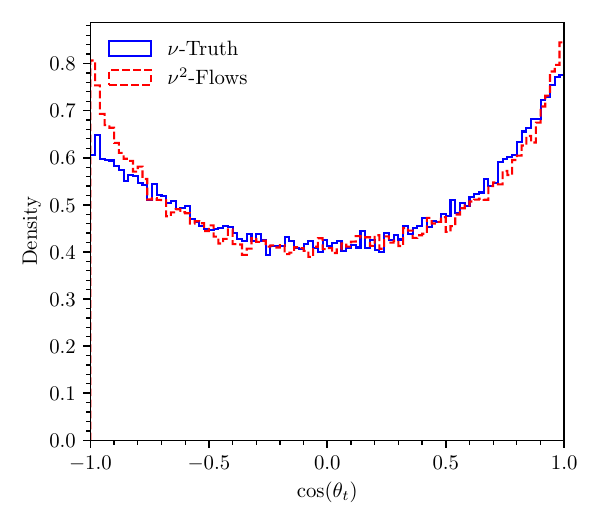}
  \caption{%
    Distributions of the $tW$ invariant mass $m_{tW}$ (top) and the
    laboratory-frame polar angle $\cos\theta_t$ of the reconstructed top
    quark (bottom) at truth and reconstructed level.
  }
  \label{fig:mtw_costheta}
\end{figure*}
The network is trained and evaluated on simulated $pp \to tW^- \to b\ell^+\ell^-\nu_\ell\bar{\nu}_\ell$
events at $\sqrt{s} = 13~\text{TeV}$. Hard-scattering matrix-element events are generated with \textsc{MadGraph5}~\cite{Alwall:2014hca}, using the NNPDF2.3 LO parton distribution functions~\cite{Ball:2012cx}, interfaced with \textsc{Pythia~8}~\cite{Bierlich:2022pfr} for parton showering and hadronization, and passed through \textsc{Delphes}~\cite{deFavereau:2013fsa} for a parametric simulation of detector response. Events are required to contain at least two oppositely charged leptons (electrons or muons) with transverse momenta $p_T > 10~\text{GeV}$ and pseudorapidity $|\eta| < 2.5$, and at least one b-jet with $p_T > 20~\text{GeV}$ and $|\eta| < 2.5$. The b-tagging is done using the loose working point in tagging algorithm available in \textsc{Delphes} which has an efficiency of around $70\%$. Events are further required to satisfy $\slashed{E}_T \ge 20$ GeV and $\Delta R_{\ell\,j} \ge 0.4$, where $\Delta R = \sqrt{\Delta\eta^2 + \Delta\phi^2}$ is the angular distance. The true neutrino information for the selected events are obtained from the event record stored in \textsc{Delphes} object.

The full simulated sample after event selection comprises $\sim 10^6$ events passing the event selection described above. This sample is divided into a training--test set in a 80:20 ratio, and $50\%$ of test samples are reserved for validation.  The network contains $2.7\times 10^6$ trainable parameters and is optimized with the \textsc{AdamW} optimizer~\cite{adamw} with a learning rate cycling from $10^{-8}$ to $10^{-3}$ and back every $50$ epochs. We use weight decay with a strength of $10^{-4}$.  Training is performed entirely on a CPU environment (Intel Core i5-14500) with a maximum of 200 epochs and early stopping with a patience of 20 epochs, monitoring the validation loss. The training of one epoch converged in approximately $1.5$ hours of wall-clock. At inference time, the trained flow is used to draw multiple samples of the neutrino momenta from the learned conditional distribution. For our analysis, we obtain a single representative neutrino-momentum prediction for each event by taking the component-wise median of the sampled momenta.

The goodness of the $\nu^2$-Flows based reconstruction is first assessed through the kinematic and angular observables that most directly control the spin-correlation analysis. Fig.~\ref{fig:nu_angular} compares the truth and reconstructed distributions of the neutrino and antineutrino pseudorapidities $\eta$ and azimuthal angles $\phi$, together with the pair-level
separations $\Delta\eta \equiv \eta_\nu - \eta_{\bar\nu}$ and $\Delta\phi \equiv \phi_\nu - \phi_{\bar\nu}$. The single-particle distributions and $\Delta\eta$ are well reproduced across their full range. The $\Delta\phi$ distribution, which encodes the spin-correlated $tW$ decay topology, is described correctly in the bulk, while the extreme tails show a modest increment in the reconstructed sample relative to truth. The $tW$ invariant mass $m_{tW}$ and the lab-frame polar angle $\cos\theta_t$ of the top quark, shown in Figs.~\ref{fig:mtw_costheta}, are likewise well described in the central region, with a small discrepancy concentrated at the edges of the $\cos\theta_t$ range. 

A quantitative summary of the reconstruction performance, in terms  of the RMS deviation, residual bias, and correlation between predicted and true momenta, is collected in
Table~\ref{tab:reconstruction_performance}. Across all components the bias remains below $1$~GeV, confirming the absence of systematic offsets, while the correlation coefficients lie in the range $0.41$--$0.57$. The larger RMS associated with the longitudinal component, $p_z \sim 155$~GeV compared with $p_{x,y} \sim 50$~GeV, reflects the intrinsic difficulty of
reconstructing the neutrino momentum along the beam axis, where the detector provides no direct constraint. A detailed assessment of how these reconstruction residuals, including the tail mismatches noted above, propagate into the downstream quantum observables is presented in later sections.

\begin{table}[!htb]
  \centering
  \caption{Reconstruction performance of the $\nu^2$-Flows model for neutrino and anti-neutrino momentum components. The root mean square (RMS), bias, and correlation coefficient between predicted and true momenta are reported for each Cartesian component.}
  \label{tab:reconstruction_performance}
  \renewcommand{\arraystretch}{1.3}
  \begin{tabular}{ccccccc}
    \hline
    \multirow{2}{*}{Statistics} & \multicolumn{3}{c}{Neutrino} & \multicolumn{3}{c}{Anti-Neutrino} \\
    \cline{2-7} 
    & $p_x^\nu$ & $p_y^\nu$ & $p_z^\nu$ & $p_x^{\bar{\nu}}$ & $p_y^{\bar{\nu}}$ & $p_z^{\bar{\nu}}$ \\
    \hline
    RMS [GeV] & $49.5$ & $50.3$ & $153.1$ & $50.0$ & $51.5$ & $155.1$ \\
    Bias [GeV] & $-0.24$ & $-0.05$ & $-0.02$ & $-0.54$ & $0.34$ & $-0.95$ \\
    Corr & $0.57$ & $0.56$ & $0.44$ & $0.57$ & $0.55$ & $0.41$ \\
    \hline
  \end{tabular}
\end{table}

\section{Quantum Correlations in the \texorpdfstring{$tW$}{tW} System}
\label{sec:GQD_tW}

Quantum discord (QD) provides a measure of non-classical correlations beyond entanglement. For a general bipartite state $\rho_{AB}$, the quantum discord with measurements on subsystem $A$ is defined as
\begin{align}
    D_A(\rho_{AB}) = I(\rho_{AB}) - \max_{\{\Pi_k^A\}} J(\rho_{AB}),
\end{align}
where $I(\rho_{AB}) = S(\rho_A) + S(\rho_B) - S(\rho_{AB})$ is the quantum mutual information, and $J(\rho_{AB}) = S(\rho_B) - \sum_k p_k S(\rho_B^k)$ is the classical correlation obtained by a projective measurement $\{\Pi_k^A\}$ on subsystem $A$; $S(\rho)=-\mathrm{Tr}[\rho\log_2 \rho]$ is the von Neumann entropy.  

For two-qubit systems, an analytic expression for quantum discord exists~\cite{Girolami:2011pzn}. In particular, for the spin density matrix of $t\bar{t}$ production, the quantum discord can be written as~\cite{Han:2024ugl}  
\begin{align}
\label{eq:ttbar_disc}
    D_A(\rho_{AB}) = 1 + \sum_i e_i \log_2 e_i 
    - \frac{1}{2} (1+\Lambda_{\rm max}) \log_2 \frac{1+\Lambda_{\rm max}}{2} 
    - \frac{1}{2} (1-\Lambda_{\rm max}) \log_2 \frac{1-\Lambda_{\rm max}}{2},
\end{align}
where $\{e_i\}$ are the eigenvalues of $\rho_{AB}$, and $\Lambda_{\rm max}$ is the largest singular value of the correlation matrix $C_{ij}$. This value corresponds to the maximal projection of the Bloch vector along the optimal measurement direction. 

For asymmetric or higher-dimensional bipartite systems, such as the qubit--qutrit $tW$ system, direct evaluation of entropic QD is computationally challenging due to the required optimization over local measurements. In these cases, \emph{geometric quantum discord} (GQD)~\cite{Dakic:2010xfz} provides a convenient alternative. GQD quantifies the minimal Hilbert--Schmidt distance between a given state and the set of classical--quantum states, defined as
\begin{align}
    D_g(\rho) = \mathrm{min}_{\chi \in \Omega}\,\mathrm{Tr}[(\rho-\chi)^2].
\end{align}
Here $\Omega$ represents the set of all states with vanishing quantum discord. For an arbitrary two-qudit state $\rho$ on $\mathcal{H}_{d_1}\otimes\mathcal{H}_{d_2}$, the exact analytical value of the GQD is given by~\cite{Loubenets:2024opb}
\begin{align}
    \label{eq:gqd-general}
    D_g(\rho) &= \frac{d_2-1}{d_1 d_2}\|r_2\|^2 + \frac14\mathrm{Tr}[T_\rho^\dagger T_\rho] - \sum_{n=1}^{d_2-1}\eta_n \nonumber\\&= \sum_{n=d_2}^{d_2^2-1}\eta_n,
\end{align}
where $\eta_1\geq \eta_2 \geq \dots \geq \eta_{d_2^2-1}\geq 0$ are the eigenvalues, listed in decreasing order, of the positive Hermitian operator
\begin{align}
    G(\rho) = \frac{d_2-1}{d_1 d_2}\ket{r_2}\bra{r_2} + \frac14 T_\rho^\dagger T_\rho,
\end{align}
$r_2$ is the Bloch vector of the subsystem of dimension $d_2$ against which the classical basis in $\Omega$ is defined, and $T_\rho$ is the correlation matrix of $\rho$. Since this result holds for arbitrary $d_1,d_2\geq 2$, it provides an exact closed form for the GQD with respect to either subsystem of the qubit--qutrit $tW$ state, not merely a bound.

For the top-quark subsystem ($d_2=2$, $d_1=3$), only the single largest eigenvalue of the resulting $3\times3$ operator is subtracted,
\begin{align}
    \label{eq:gqd-top}
    D_g^{t}(\rho_{tW}) = \frac16\|\vec{P}_t\|^2 + \frac14\mathrm{Tr}[C^\dagger C] - \eta_1^t = \eta_2^t + \eta_3^t,
\end{align}
while for the $W$-boson subsystem ($d_2=3$, $d_1=2$), the top two eigenvalues of the resulting $8\times8$ operator are subtracted,
\begin{align}
    \label{eq:gqd-w}
    D_g^{W}(\rho_{tW}) &= \frac13\|\vec{B}_W\|^2 + \frac14\mathrm{Tr}[C^\dagger C] - \left(\eta_1^W + \eta_2^W\right) \nonumber \\& = \sum_{n=3}^{8}\eta_n^W,
\end{align}
where $\vec{P}_t$ and $\vec{B}_W \in \{\vec{P}_W, T_W\}$ are the three- and eight-dimensional Bloch vectors of the top quark and $W$ boson, respectively, $C$ is the spin-correlation matrix defined via $P P^{tW}_{ij}$ and $P T^{tW}_{ijk}$ in Eq.~\eqref{eq:rho_tw}, and $\{\eta_n^{t}\}$, $\{\eta_n^{W}\}$ are the eigenvalues, in decreasing order, of
\begin{align}
    G^t(\rho_{tW}) &= \frac16 \ket{\vec{P}_t}\bra{\vec{P}_t} + \frac14 C^\dagger C, \nonumber\\
    G^W(\rho_{tW}) &= \frac13 \ket{\vec{B}_W}\bra{\vec{B}_W} + \frac14 C^\dagger C.
\end{align}
This formulation enables efficient, exact computation of non-classical correlations in mixed and asymmetric systems like $tW$.

Further, to explicitly verify the presence of entanglement, we also compute the negativity based on the Peres--Horodecki partial transpose (PPT) criterion~\cite{Peres:1996dw,Horodecki:1996nc,Vidal:2002zz}. For a bipartite state $\rho_{tW}$, the negativity is defined as
\begin{align}
    \label{eq:ppt}
    \mathcal{C}^{\mathrm{PPT}}[\rho_{tW}] = \frac{\|\rho^{T_{t/W}}\|_1 - 1}{2} = \sum_i |\lambda_i^-(\rho^{T_{t/W}})|,
\end{align}
where $\rho^{T_{t/W}}$ denotes the partial transpose with respect to subsystem $t/W$, $\|\cdot\|_1$ is the trace norm, and $\lambda_i^-$ are the negative eigenvalues of $\rho^{T_{t/W}}$. While non-zero negativity signals entanglement, GQD captures more general quantum correlations even in separable states.

\begin{figure*}[!htb]
    \centering
    \includegraphics[width=0.99\textwidth]{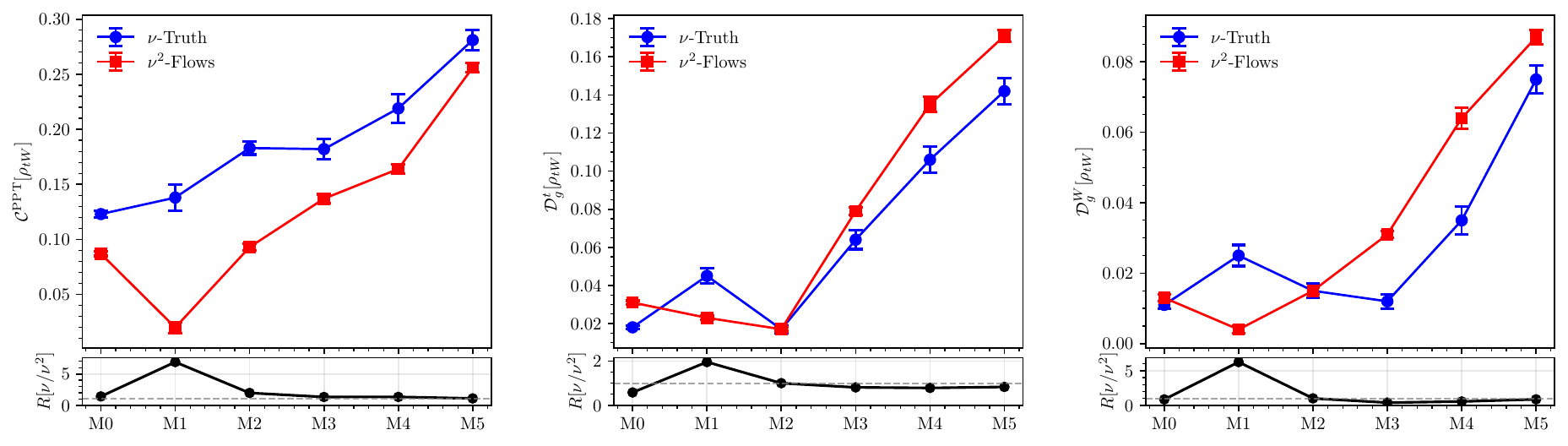}
    \caption{Standard Model predictions for the PPT-based entanglement measure (right panel) and geometric quantum discord w.r.t to top quark (middle panel) and $W$ boson (left panel) as functions of $m_{tW}$ at $\nu-$truth and reconstruction level.}
    \label{fig:ppt_gqd}
\end{figure*}

Further, since the closed analytic forms of $\mathcal{C}^{\mathrm{PPT}}$ and $D_g^{t/W}$ for the mixed $tW$ system in terms of the polarization and spin correlation parameters are not known, the uncertainties propagated to the final QIs from the elements of the density matrix are non-trivial to obtain. To estimate the errors on the QIs, we therefore selected a random $40\%$ subset of events from the SM datasets and computed the QIs. This procedure was repeated 1000 times, and the resulting QI values were fitted to a normal distribution. From this fit, the standard deviation was taken as the $1\sigma$ error on the estimation of the QIs. The corresponding estimate of QIs at the SM point is listed in Table~\ref{tab:sment} for unbinned along with five disjoint region of $m_{tW}$ at $\nu-$truth and reconstruction level. The graphical representation of the comparison are also shown in Fig.~\ref{fig:ppt_gqd}.

\begin{table}[!htb]
    \centering
    \caption{Negativity and geometric quantum discord along with $1\sigma$ standard deviation for the $tW$ process at the level of $\nu-$truth and reconstruction at five different disjoint region of invariant mass of $tW$ system.}
    \label{tab:sment}
    \scriptsize
    \renewcommand{\arraystretch}{1.4}
    \begin{tabular*}{0.99\textwidth}{@{\extracolsep{\fill}}ccccccc@{}}
    \hline
    \multirow{2}{*}{Region} & \multicolumn{3}{c}{True} & \multicolumn{3}{c}{Reconstructed} \\
    \cline{2-7} 
         & $\mathcal{C}^{\mathrm{PPT}}[\rho_{tW}]$ & $D_g^t(\rho_{tW})$ & $D_g^W(\rho_{tW})$ & $\mathcal{C}^{\mathrm{PPT}}[\rho_{tW}]$ & $D_g^t(\rho_{tW})$ & $D_g^W(\rho_{tW})$ \\
    \hline
    M0: Unbinned &$0.123 \pm 0.003$ & $0.018\pm0.001$ & $0.011\pm 0.001$ & $0.087\pm0.002$ & $0.031\pm 0.001$ & $0.013\pm 0.001$  \\
    M1: $m_{tW} \leq 300~\mathrm{GeV}$  & $0.138\pm 0.012$ &$0.045\pm0.004$  & $0.025\pm0.003$ & $0.020\pm0.005$ &$0.023\pm0.001$ & $0.004\pm0.001$  \\
    M2: $300 < m_{tW} \le 400$ GeV & $0.183\pm 0.006$ & $0.017\pm0.002$& $0.015\pm 0.002$ & $0.093\pm 0.003$ & $0.017\pm0.001$& $0.015\pm 0.001$  \\
    M3: $400< m_{tW} \le 500$ GeV & $0.182\pm 0.009$&$0.064\pm0.005$ & $0.012\pm 0.002$ & $0.137\pm 0.004$ &$0.079\pm0.002$ &$0.031\pm 0.001$ \\
    M4: $500 < m_{tW} \le 600$ GeV & $0.219\pm 0.013$ &$0.106\pm0.007$ & $0.035\pm 0.004$ & $0.164\pm 0.004$ &$0.135\pm0.004$ &$0.064\pm 0.003$ \\
    M5: $m_{tW} > 600$ GeV & $0.281\pm 0.009$ & $0.142\pm0.007$& $0.075\pm 0.004$ & $0.256\pm 0.004$ &$0.171\pm0.003$ &$0.087 \pm 0.002$ \\
    \hline
    \end{tabular*}
\end{table}

Both correlation measures exhibit an overall increasing trend with $m_{tW}$, with some non-monotonicity in the intermediate bins. At truth level, $\mathcal{C}^{\mathrm{PPT}}$ rises from $0.138$ in M1 to $0.281$ in M5, while the reconstructed value grows more steeply, from $0.020$ to $0.256$. The geometric discord $\mathcal{D}_g^t$ increases from $0.045$ to $0.142$ at truth level, with a shallow dip at M2, and from $0.023$ to $0.171$ after reconstruction. The $W$-boson discord $\mathcal{D}_g^W$ rises from $0.025$ to $0.075$ at truth level, exhibiting a shallow minimum around M2--M3, and from $0.004$ to $0.087$ after reconstruction. The non-zero values of all three observables across every mass bin confirm that non-classical correlations persist in the $tW$ system over the full $m_{tW}$ range probed. It is further confirmed by computing the QIs for the full integrated phase space (M0) and all QIs are non-zero significantly above $1\sigma$ statistical error.

The three observables respond differently to the $\nu^2$-Flows reconstruction. For $\mathcal{C}^{\mathrm{PPT}}$, the reconstructed values are systematically lower than the truth in every bin, with the ratio $R \equiv \mathcal{O}_{\mathrm{truth}}/\mathcal{O}_{\mathrm{reco}}$ decreasing monotonically from $\sim 7$ in M1 to $\sim 1.1$ in M5. The neutrino-momentum smearing therefore suppresses partial-transpose negativity throughout the probed range. The discord observables behave differently. At low $m_{tW}$, $\mathcal{D}_g^t$ is underestimated by the reconstruction ($R \simeq 2$ in M1), but the ordering reverses from M3 onward, where the reconstructed values exceed the truth ($R \simeq 0.5$--$0.8$). A similar pattern is seen for $\mathcal{D}_g^W$, with a much larger low-mass bias ($R \simeq 6$ in M1) and improvement at larger $m_{tW}$.

The reconstruction fidelity improves markedly with $m_{tW}$. At M5 ($m_{tW} \gtrsim 600~\mathrm{GeV}$), the fractional bias $|1-R|$ is at most $\sim 10\%$ for $\mathcal{C}^{\mathrm{PPT}}$ and $\sim 14-17\%$ for the two discord observables, compared with factors of $2$--$7$ in M1. This is also the regime where higher-dimensional SMEFT operators generate their largest deviations from the SM predictions, scaling as $m_{tW}^2/\Lambda^2$, so the reconstruction is most reliable precisely where the new-physics sensitivity is greatest.

Next, we estimate the effect on the quantum observables with the inclusion of reducible backgrounds, viz.\ $t\bar{t}$ and $WWj$.

\section{Inclusion of Backgrounds}
\label{sec:background}

We consider two major background processes viz. $t\bar{t}$ and $W^+W^-j$ production, at leading order. Since the signal and background events are induced by different production channels but pass the same event selection, the observed sample is characterized by a mixture of density matrices belonging to each process. Following the recipe of Ref.~\cite{Fabbrichesi:2023jep}, we write the density matrix for the selected $tW$-tagged sample as
\begin{align}
    \label{eq:rho_bkg}
    \rho_{tW}^\prime = \alpha_{tW}\,\rho_{tW} + \alpha_{t\bar{t}}\,\rho_{t\bar{t}} + \alpha_{WWj}\,\rho_{WWj},
\end{align}
with $\sum_i \alpha_i = 1$. Here $\alpha_i$ is the fraction of events tagged as $tW$ that are truly of process $i$, generalizing the signal-to-background ratio $S/(S+B)$ to three classes.

The signal and both backgrounds are generated at leading order at $\sqrt{s}=13$ TeV using \textsc{MG5} with the 5-flavor scheme, and pass the selection criteria discussed in Sec.~\ref{sec:reco}. Writing $N_i$ for the expected number of events of process $i$ passing this selection, and $\epsilon_{i\to tW}\equiv P(\mathrm{tagged}=tW\mid\mathrm{true}=i)$ for the probability that a selected event of true process $i$ is tagged as $tW$ by the classifier described below, the cross-section-weighted mixture fractions are
\begin{align}
    \label{eq:alpha}
    \alpha_i = \frac{N_i\,\epsilon_{i\to tW}}{\displaystyle\sum_{j\in\{tW,\,t\bar t,\,WWj\}} N_j\,\epsilon_{j\to tW}}, \qquad N_i = \mathcal{L}\,\sigma_i\,\epsilon^d_i,
\end{align}
where $\mathcal{L}$ is the integrated luminosity, $\sigma_i \equiv \sigma_i^{(0)}\times\Gamma_i$ is the process cross section at the decay level with $\sigma_i^{(0)}$ the Born-order production cross section and $\Gamma_i$ the branching fraction into the dileptonic final state, and $\epsilon_i^d$ is the kinematic acceptance of the selection cuts at the detector level.

The leading-order cross sections obtained from this generation are
\begin{align}
    \label{eq:xsec}
    \sigma_{tW} &= 2.562 \pm 0.003~\mathrm{pb},\nonumber\\
    \sigma_{t\bar{t}} &= 22.73\pm0.014~\mathrm{pb},\\
    \sigma_{WWj} &= 1.902\pm 0.002~\mathrm{pb},\nonumber
\end{align}
with the corresponding kinematic acceptances of the selection cuts of Sec.~\ref{sec:reco},
\begin{align}
    \label{eq:det_eff}
    \epsilon_{tW}^d &= 0.21,\nonumber \\
    \epsilon_{t\bar t}^d &= 0.17, \\
    \epsilon_{WWj}^d &= 0.02.\nonumber
\end{align}

To determine $\epsilon_{i\to tW}$, we extract a list of features from the selected events at the reconstruction level, listed in Table~\ref{tab:features}, and train a classifier implemented with \texttt{XGBoost}~\cite{Chen_2016}, a gradient-boosted decision tree (GBDT) algorithm that builds an ensemble of weak learners in a stage-wise fashion.

\begin{table}[!t]
    \centering
    \renewcommand{\arraystretch}{1.3}
    \caption{List of the features used in training the boosted decision trees employed for classification of $tW$ events.}
    \label{tab:features}
    \begin{tabular}{cc}
    \hline
     Features & Description \\
     \hline
     $p_i,\,i\in\{\ell,\mathrm{jets}\}$ & Four momenta of visible particles\\
      $p_T^i$ & Transverse momenta of visible particles\\
      $\phi_i,\,\eta_i$ & Azimuth and pseudorapidity\\
      $\Delta R_{ij}$ & Angular distance \\
      $m_{ij}$ & Pair invariant mass \\
      $p_T^{(ij)}$ & Transverse momenta of pairs \\
      $H_T$ & Scalar sum of transverse momenta\\
      $\slashed{E}_T$ & Missing transverse momentum\\
      $m_{\ell^-\ell^+j}$ & Invariant mass of visible particles\\
      $m_W^T$ & $\equiv \sqrt{2\,p_T^\ell\,p_T^\nu\,\Big(1-\cos(\Delta\phi_{\ell\nu})\Big)}$ \\
      $n_j$ & Jet multiplicity \\
      \hline
    \end{tabular}
\end{table}

The input features are rescaled to the unit interval using a min--max normalization fitted on the training set,
\begin{equation}
    \tilde{x}_j = \frac{x_j - \min_i x_{ij}}{\max_i x_{ij} - \min_i x_{ij}},
    \label{eq:minmax}
\end{equation}
where the minimum and maximum are computed over the training events $i$ for each feature $j$; the same transformation, with the training-set extrema, is applied to the validation and test sets. We take a total of $10^6$ events, split into $80\%$ training and $20\%$ testing samples, with the latter further divided into validation and test subsets at a $50{:}50$ ratio.

The classifier outputs class probabilities over three classes ($tW$, $t\bar{t}$, $WWj$) through a softmax,
\begin{equation}
    p_{ic} = \frac{e^{z_{ic}}}{\sum_{c^\prime=1}^{C} e^{z_{ic^\prime}}},
    \label{eq:softmax}
\end{equation}
and is trained by minimizing the categorical cross-entropy
\begin{equation}
    \mathcal{L}_{\mathrm{loss}} = -\frac{1}{N}\sum_{i=1}^{N}\sum_{c=1}^{C} y_{ic}\,\log p_{ic},
    \label{eq:mlogloss}
\end{equation}
where $N$ is the number of events, $C=3$ is the number of classes, $z_{ic}$ are the raw scores of the boosted ensemble, $y_{ic}$ is the one-hot encoded true label of event $i$ for class $c$, and $p_{ic}$ is the corresponding predicted class probability. We determine the classifier hyperparameters through a randomized search over tree depth, learning rate, sub-sampling fractions, and regularization strengths. The optimal configuration uses a maximum tree depth of $8$, a learning rate of $\eta = 0.01$, and $2000$ boosting rounds, with row and column sub-sampling fractions of $0.7$ and $0.9$, respectively. Regularization is enforced through an $L_2$ penalty $\lambda = 1.0$, while the $L_1$ penalty is set to zero, and a minimum loss reduction of $\gamma = 0.5$ is required for a further split.

\begin{figure}[!htb]
    \centering
    \includegraphics[width=0.75\textwidth]{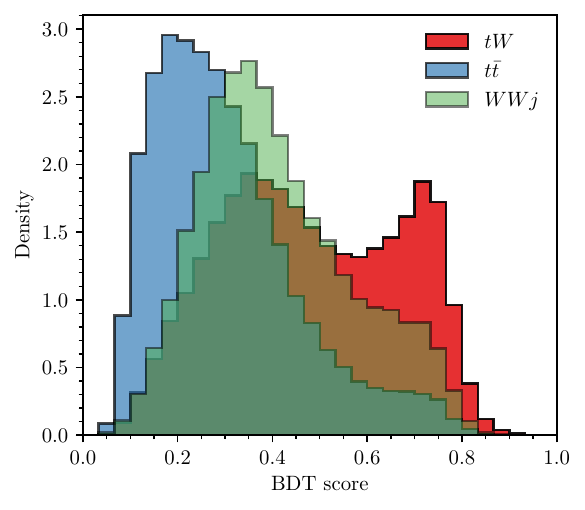}
    \caption{Boosted decision tree output score distribution for signal $tW$ and two reducible backgrounds viz. $t\bar{t}$ and $WWj$ processes .}
    \label{fig:bdt_score}
\end{figure}

To maximize the purity (see BDT score distribution in Fig.~\ref{fig:bdt_score}), we choose a classifier threshold of $0.85$ at which the tagging efficiencies $\epsilon_{i\to tW}$ are obtained as,
\begin{align}
   \label{eq:mistag}
    \epsilon_{tW\to tW} &= 0.33\,\%,\nonumber\\
    \epsilon_{t\bar{t} \to tW} &= 0.02\,\%,\\
    \epsilon_{WWj\to tW} &= 0.03\,\%.\nonumber
\end{align}
These probabilities corresponds to around $249$ $tW$ events at $\mathcal{L}=140$ fb$^{-1}$ and similarly $108$ and $2$ events for $t\bar t$ and $WWj$ process, respectively at the leading order. Combining Eqs.~\eqref{eq:xsec}--\eqref{eq:mistag} in Eq.~\eqref{eq:alpha} gives the mixture fractions of the $tW$-tagged sample,
\begin{align}
    \label{eq:alpha_final}
    \alpha_{tW} &\simeq 0.69,\nonumber\\
    \alpha_{t\bar{t}} &\simeq 0.30,\\
    \alpha_{WWj} &\simeq 0.01.\nonumber
\end{align}
Despite the low probability of $t\bar{t}$ to be tagged as $tW$ events, the overall contribution to the $\rho_{tW}^\prime$ from top pair events stands around $30\%$ owing to large cross~section i.e. $\mathcal{O}(1)$ greater than that of $tW$ at the leading order. Although the tagging performance could, in principle, be further improved with a more expressive and robust classifier, the primary objective of this study is to quantify the impact of background contamination on the quantum observables. Consequently, further optimization of the tagger is not pursued in this work.

The tagged events undergoes the neutrino reconstruction (as discussed above) and further extraction of the elements of the density matrix is done. In Table~\ref{tab:bkgent}, we list down the central values of the three quantum observables along with the $1\sigma$ error for $\rho_{tW}^\prime$ defined in Eq.~\eqref{eq:rho_bkg} following the same iterative procedure discussed above.

\begin{table}[!htb]
    \centering
     \caption{\label{tab:bkgent}Negativity and geometric quantum discord along with $1\sigma$ standard deviation for $\rho_{tW}^\prime$ defined in Eq.~\eqref{eq:rho_bkg} at the level of reconstruction for five different disjoint region of invariant mass of $tW$ system.}
    \renewcommand{\arraystretch}{1.4}
    \begin{tabular*}{0.98\textwidth}{@{\extracolsep{\fill}}cccc@{}}
    \hline
      $m_{tW}$ & $\mathcal{C}^{\mathrm{PPT}}$ & $D_g^t$ & $D_g^W$  \\
      \hline
       M0 & $0.075\pm 0.001$ & $0.029\pm 0.001$ & $0.013\pm 0.001$ \\
       M1 & $0.019\pm0.005$ & $0.019\pm0.002$ & $0.005\pm 0.001$  \\
       M2 & $0.067\pm0.003$ & $0.013\pm0.001$ & $0.012\pm0.001$ \\
       M3 & $0.117\pm0.003$ & $0.066\pm0.002$ & $0.028\pm0.001$ \\
       M4 & $0.146\pm0.003$ & $0.112\pm0.003$ & $0.053\pm0.002$ \\
       M5 & $0.235\pm0.004$ & $0.150\pm0.003$ & $0.079\pm0.002$ \\
      \hline
    \end{tabular*}
\end{table}
Comparing the reconstructed entries of Tables~\ref{tab:sment} and~\ref{tab:bkgent}, the $tW$ background
uniformly dilutes all three correlation measures. The background-inclusive values lie below the signal-only ones in every mass region, the only exception being $D_g^W$ in M1, where the two agree within uncertainties. The magnitude of the dilution is observable-dependent. Negativity is the most sensitive, with $\mathcal{C}^{\mathrm{PPT}}$ reduced by up to $\sim 30\%$ in the
intermediate-mass bins and by $\sim 5$--$15\%$ elsewhere, whereas the discord-type measures are more robust: across M1--M5, $D_g^t$ is diluted by $\sim 10$--$25\%$ and $D_g^W$ by $\sim 10$--$20\%$. In the
high-mass tail (M3--M5), where the signal purity is largest, all three observables settle to a common $\sim 10$--$15\%$ dilution, indicating that the background predominantly degrades the low-mass, low-purity  region.

\section{Quantum correlation in presence of anomalous dipole moments}
\label{sec:anom}
\begin{figure*}[!htb]
    \centering
    \includegraphics[width=0.99\textwidth, height=11cm]{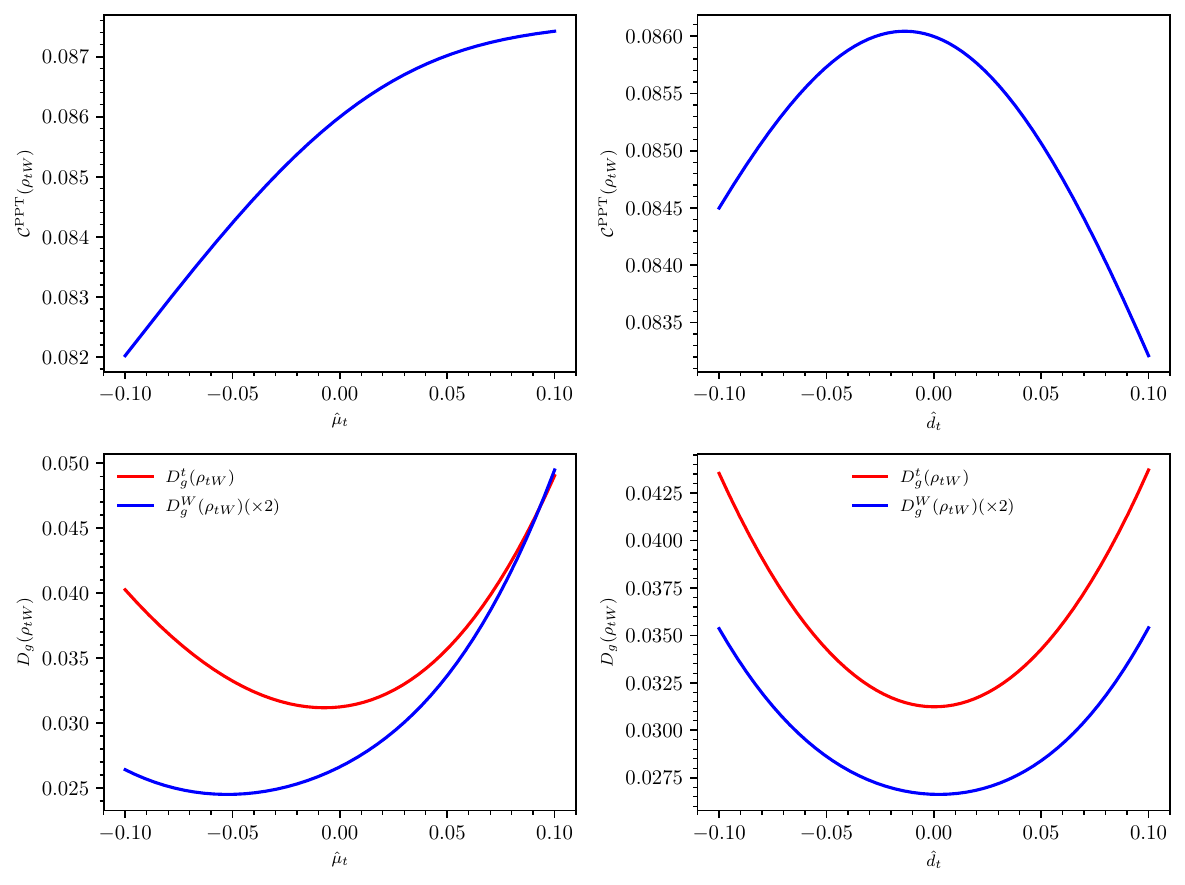}
    \caption{Distribution of negativity and geometric quantum discord for $tW$ system in presence of anomalous CEMDM of top quark at the reconstruction level. The distribution for $D_g^W$ are rescaled by a factor of two for visual comparison.}
    \label{fig:1danom}
\end{figure*}
Beside being a relativistic platform to test non-classical correlations in mixed dimensional bipartite state, the associated $tW$ process also offers a unique place to study the anomalous chormomagnetic dipole moment~(CMDM) and chromoelectric dipole moment~(CEDM) of the top quark. These moments are produced due to the presence of $gt\bar{t}$ couplings, and within the SM at leading order, both the moments are zero. In the SM, the next-to-leading order~(NLO) contribution from QCD and electroweak~(EW) to CMDM were found to be $0.056$~\cite{Martinez:2007qf}. CEDM only appears at three loop levels, arising from the complex phase in the Cabibbo-Kobayashi-Maskawa~(CKM) matrix~\cite{Czarnecki:1997bu,Khriplovich:1985jr}. 

The $gt\bar{t}$ Lagrangian, including the SM and the anomalous contribution can be parametrized in terms of effective Lagrangian as~\cite{Haberl:1995ek,CMS:2019kzp}:
	\begin{align}
		\label{eq:Laggtt}
			\mathscr{L}_{gtt} = -g_s\bar{t}\gamma^\mu G^a_\mu \frac{\lambda^a}{2}t -\frac{g_s}{2m_t}\bar{t}\sigma^{\mu\nu}\left(\hat{\mu}_t+i\hat{d}_t\gamma_5\right)G^a_{\mu\nu}\frac{\lambda^a}{2}t.
		\end{align}
Here $\hat{\mu}_t$ and $\hat{d}_t$ are the top quark chromomagnetic and chromoelectric dipole moments, respectively, $\lambda$ are the Gell-Mann matrices, $G_{\mu\nu}^a$ the gluon field strength tensor, $g_s$ the strong coupling constant and $m_t$ the top quark mass. The gluonic field tensor is defined as 
\begin{align}
    G_{\mu\nu}^a = \partial_\mu G^a_\nu - \partial_\nu G^a_\nu - g_sf_{abc}G^b_\mu G^c_\nu,
\end{align}
where $g_s$ is the strong coupling constant and $f^{abc}$ is the QCD structure constant.

The second part of Eq.~(\ref{eq:Laggtt}) can arise from the dimension-six operator~\cite{Buchmuller:1985jz,Aguilar-Saavedra:2018ggp}
    \begin{equation}
		\mathscr{O}_{uG\Phi} =\left(\bar{t}\lambda^a\sigma^{\mu\nu}t\right)\tilde{\Phi}G^a_{\mu\nu},
	\end{equation}  
	where $\Phi$ is the Higgs doublet and $\tilde{\Phi} = i\tau_2\Phi^\star$. The couplings $\mu_t$ and $d_t$ are $CP$-even and odd respectively and they are related to operator $\mathscr{O}_{uG\Phi}$ as:
		\begin{align}
			\hat{d}_t &= \frac{1}{\sqrt{2}\,m_t}\text{Im}\left[C_{uG\Phi}\right]\frac{v\,m_t}{\Lambda^2} \nonumber\\
			\hat{\mu}_t &= \frac{1}{\sqrt{2}\,m_t}\text{Re}\left[C_{uG\Phi}\right]\frac{v\,m_t}{\Lambda^2},
		\end{align}
	where $v\sim 246$ GeV is the vacuum expectation value~(VEV) of the Higgs field and $\Lambda$ is the characteristic new physics scale.
The current best experimental probe on CMDM and CEDM of top quark are reported by CMS~\cite{CMS:2019kzp},
\begin{align}
\label{eq:mut}
\hat{\mu}_t = -0.024^{+0.013}_{-0.009}\,(\mathrm{stat.})^{+0.016}_{-0.011}\,(\mathrm{syst.}),
\end{align}
and 
\begin{align}
\label{eq:dt}
-0.03 \le \hat{d}_t| \le +0.03 
\end{align}
at $95\%$ confidence level (CL).


Figure~\ref{fig:1danom} shows the dependence of the Peres--Horodecki negativity and the geometric discord on the anomalous top-quark dipole moments for the unbinned case. The negativity $C^{\mathrm{PPT}}$ and the discords $D_g^t$ and $D_g^W$ (rescaled by a factor of two) exhibit opposite curvature: $C^{\mathrm{PPT}}$ is concave, with a maximum shifted away from the SM point and lies at the positive axis beyond the scanned region, while both discords are convex, with minima close to the SM point. The near-symmetric response under $\hat{\mu}_t\to-\hat{\mu}_t$ and $\hat{d}_t\to-\hat{d}_t$ in all panels indicates that the leading dependence is quadratic, with a small linear admixture visible as a slight shift of the extrema away from the SM point.

\begin{figure*}[!htb]
    \centering
    \includegraphics[width=0.33\textwidth]{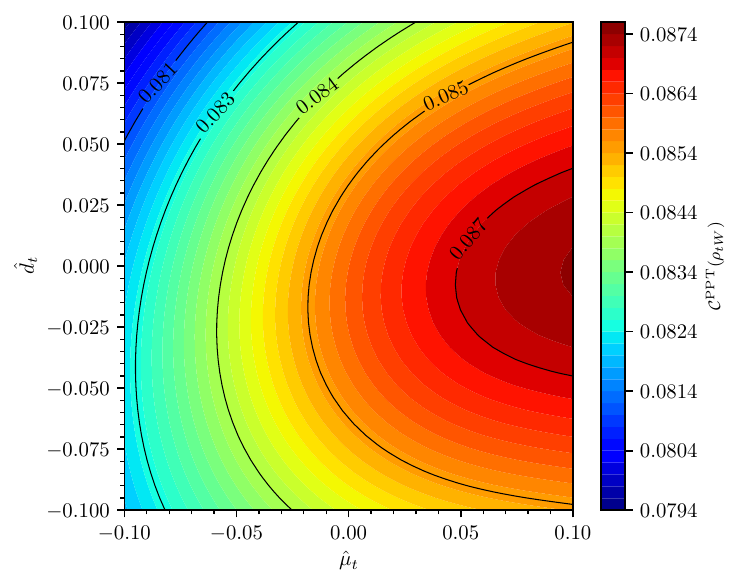}
    \includegraphics[width=0.66\textwidth]{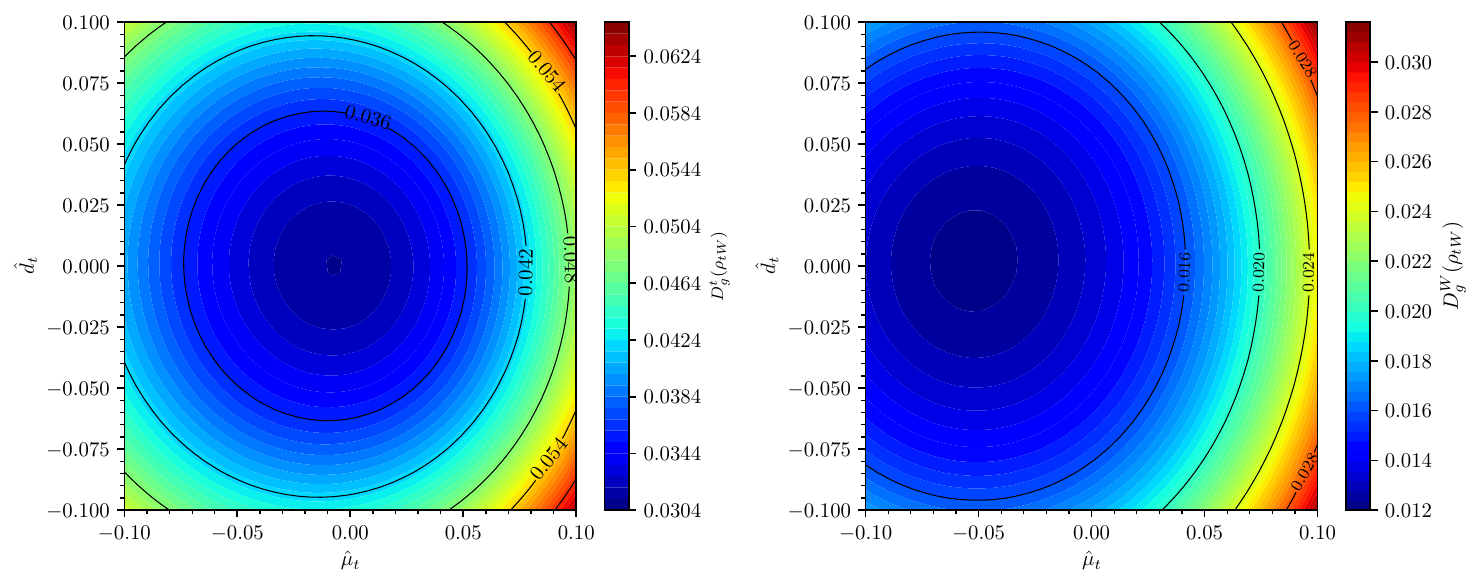}
    \caption{Two dimensional density plot for negativity $\mathcal{C}^{\mathrm{PPT}}[\rho_{tW}]$ (left panel) and geometric quantum discord $D_g^W(\rho_{tW})$ (right panel) as a function of anomalous $\mu_t$ and $d_t$. The distribution are obtained for $pp\to tW \to \ell^-\ell^+b\slashed{E}$ final state at $\sqrt{s}=13$ TeV at the reconstruction level.}
\label{fig:ppt_gqd_2d}
\end{figure*}

The same structure persists in the two-dimensional correlation distributions of Fig.~\ref{fig:ppt_gqd_2d}, which show the negativity and the top- and $W$-subsystem discords in the $(\hat{\mu}_t,\hat{d}_t)$ plane at reconstruction level. For $C^{\mathrm{PPT}}$, the distribution is dominated by an approximately circular contour centered away the SM point, a behavior dictated by the concave nature of $\hat{\mu}_t$ dependence. The $\mathcal{C}^{\mathrm{PPT}}$ decreases as one moves away from this point  on all four quadrants. The discords $D_g^t$ and $D_g^W$ show the opposite pattern, with the QIs increasing on all quadrants on moving away from the SM point.

Both the one- and two-dimensional profiles follow from the quadratic dependence of the polarization and spin-correlation coefficients that enter the $tW$ density matrix. Schematically, the quantum-information observables $\mathrm{QI}\in\{C^{\mathrm{PPT}},D_g^t,D_g^W\}$ can be approximated as
\begin{align}
\label{eq:qi_func}
\mathrm{QI}(\hat{\mu}_t,\hat{d}_t)
\sim
f_0 + f_1\hat{\mu}_t + f_2\hat{d}_t + f_3\hat{\mu}_t\hat{d}_t
+ f_4\hat{\mu}_t^2 + f_5\hat{d}_t^2 + \mathcal{O}(\text{higher order}) ,
\end{align}
where the higher-order terms originate from the nonlinear construction of the quantum-information measures from the density-matrix elements. Since $C^{\mathrm{PPT}}$ and $D_g$ are nonlinear functions of the eigenvalues and matrix elements of $\rho_{tW}$, Eq.~\eqref{eq:qi_func} is a schematic leading-order description rather than an exact polynomial expansion. The opposite signs of the curvature observed in Fig.~\ref{fig:1danom} imply that the quadratic coefficients $f_4,f_5$ are positive for the discords and negative for the negativity, with $|f_4|,|f_5|$ larger for $C^{\mathrm{PPT}}$ and $D_g^t$ than for $D_g^W$. The displaced extrema visible in the one-dimensional scans arise from the interference between the linear and quadratic terms, $\hat{\mu}_t^{\mathrm{ext}}\approx -f_1/(2f_4)$ and $\hat{d}_t^{\mathrm{ext}}\approx -f_2/(2f_5)$.  The opposite curvature of the negativity and the discords reflects their different functional dependence on the eigenvalues of $\rho_{tW}$ and $\rho_{tW}^{T}$. Negativity is dominated by the negative eigenvalue and decreases as the state becomes more mixed away from the SM point, whereas the discords measure the correlation between the subsystems and increase as the state departs from the separable SM configuration.

\begin{figure*}[!htb]
    \centering
    \includegraphics[width=0.99\textwidth]{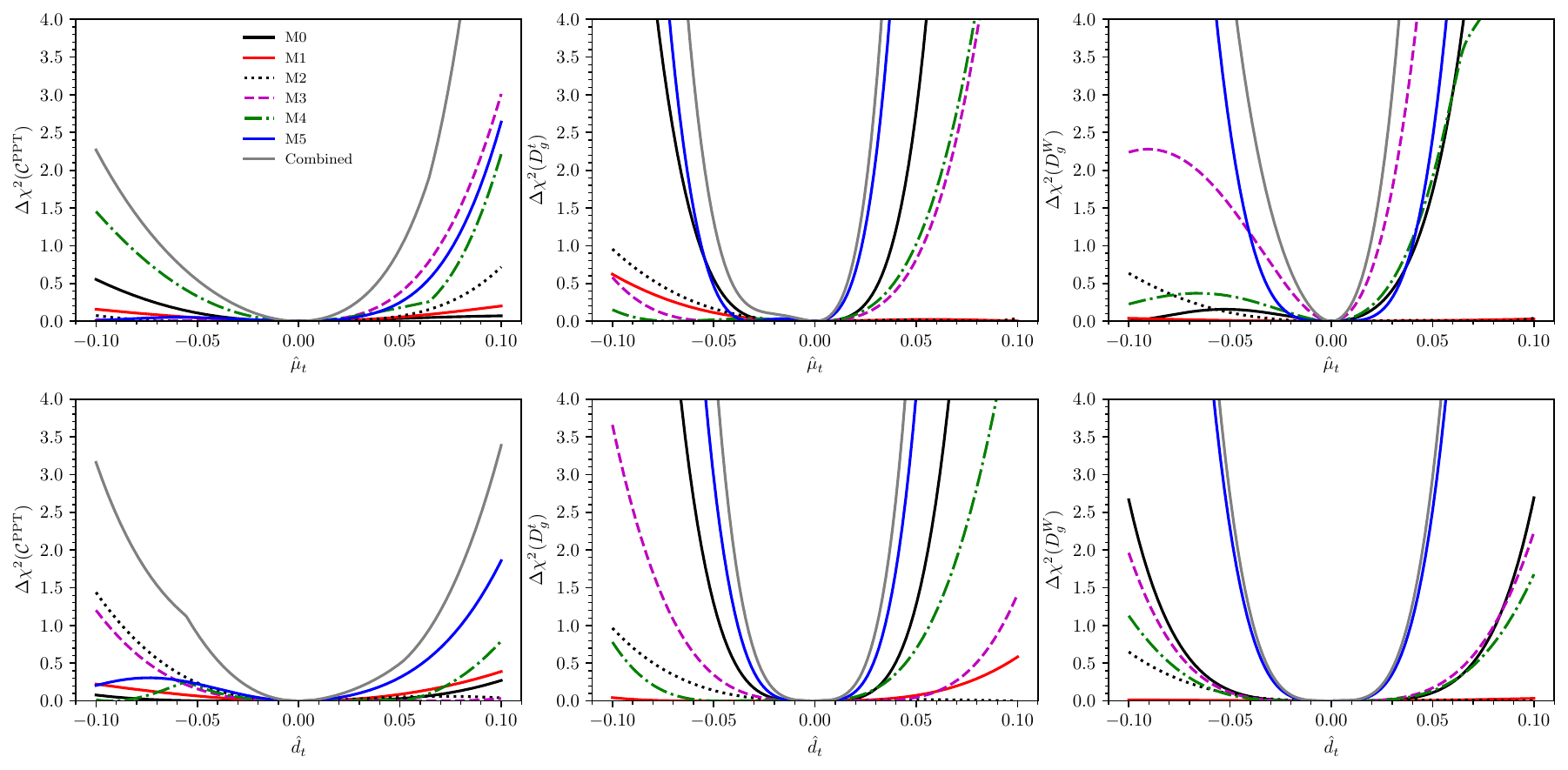}
    \caption{\label{fig:chi} Chi-squared distribution for negativity and geometric quantum discord as a function of anomalous CMDM ($\hat{\mu}_t$) (top row) and CEDM ($\hat{d}_t$) (bottom row). The distribution are obtained for  $pp\to tW \to \ell^-\ell^+b\slashed{E}$ state at the reconstruction level at an integrated luminosity of $140$ fb$^{-1}$ for different disjoint region of $m_{tW}$ defined in Tab.~\ref{tab:sment}. }
\end{figure*}

\begin{figure*}[!htb]
    \centering
    \includegraphics[width=0.99\linewidth]{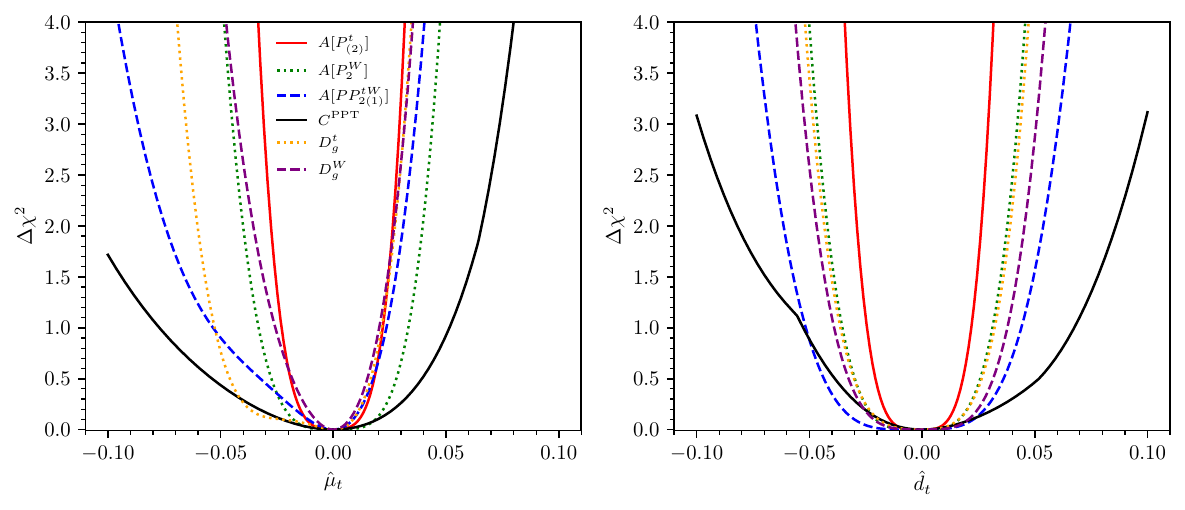}
    \caption{Values of combined $\chi^2$ from five disjoint $m_{tW}$ regions of top three sensitive asymmetries along with three quantum observables as a function of one anomalous dipole moment of top quark. The distribution are obtained for $pp\to tW \to \ell^-\ell^+ b\slashed{E}$ final state at $\sqrt{s}=13$ TeV and an integrated luminosity of $\mathcal{L}=140$ fb$^{-1}$. }
    \label{fig:compare_chi}
\end{figure*}

Next, we quantify the sensitivity of the quantum observables to the anomalous couplings via chi-squared distribution defined as
\begin{equation}
\Delta\chi^2(\hat{\mu}_t,\hat{d}_t) = \frac{\big(\mathrm{QO}(\hat{\mu}_t,\hat{d}_t)-\mathrm{QO}(0,0)\big)^2}{\delta\mathrm{QO}(0,0)^2},
\end{equation}
where $\delta\mathrm{QO}(0,0)$ is the SM estimation error. As the analytic dependence of QI on the polarization and spin-correlation asymmetries is not known in closed form, we rescale the $1\sigma$ statistical error from Table~\ref{tab:sment}, obtained at $\mathcal{L}_0 \approx 1000~\mathrm{fb}^{-1}$ (estimated from MC simulation), to the target luminosity via $\epsilon_1 = \epsilon_0\sqrt{\mathcal{L}_0/\mathcal{L}_1}$. Fig.~\ref{fig:chi} shows the resulting one-parameter $\Delta\chi^2$ profiles for $\hat{\mu}_t$ and $\hat{d}_t$ at $\mathcal{L}_1 = 140~\mathrm{fb}^{-1}$. From the Fig.~\ref{fig:chi}, we note that in the case of top-quark dipole $\hat{\mu}_t$, the geometric discord of the $W$ boson subsystem, $D_g^W$, provides the dominant sensitivity across the entire scan range, followed by $D_g^t$. The negativity $\mathcal{C}^{\mathrm{PPT}}$ remains the least sensitive observable for  both $\hat{\mu}_t$ and $\hat{d}_t$, with its $\Delta\chi^2$ curve staying roughly an order of magnitude below that of $D_g^t$ and $D_g^W$. For $\hat{d}_t$, the $D_g^t$ and $D_g^W$ provides a comparable response.

Across all observables and both couplings, the high-$m_{tW}$ bin M5 ($m_{tW} \geq 600~\mathrm{GeV}$) carries the dominant share of the combined $\Delta\chi^2$, except for $D_g^W$ dependence of $\hat{\mu}_t$, where M3 ($400 \le m_{tW} \le 500$ GeV) region provides the dominant contribution to $\chi^2$. This is consistent with the expectation that the leading SMEFT contributions scale with energy and therefore populate the tail of the $m_{tW}$ distribution, where the SM cross section is suppressed and a given absolute deviation in the quantum observables corresponds to a larger relative effect. The remaining bins M1--M4 contribute progressively less, with M1 ($m_{tW} \le 300$ GeV) giving the weakest individual constraint. The combined sensitivity (gray curve) is obtained by adding the contribution from all five region at the level of $\chi^2$. Further, the unbinned $\chi^2$ provides a competitive sensitivity to the combined case with the geometric discord for both the anomalous moments. 

We also quantify the sensitivity reach of the constituents of the density matrix i.e., polarizations and spin correlations asymmetries to the anomalous dipole moments of the top quark. The error function for the asymmetries, $A$, is given by
\begin{align}
    \label{eq:syst_err}
    \delta A = \sqrt{\frac{1-A^2}{\mathcal{L}\,\sigma} + \epsilon_A\,\sigma^2},
\end{align}
where $\sigma$ and $\epsilon_A$ denotes SM cross~section and systematic error on asymmetry measurement, respectively. For that we select top three sensitive asymmetries in the combined region, the corresponding distribution are shown in Fig.~\ref{fig:compare_chi} at $\mathcal{L}=140$~fb$^{-1}$. The polarization asymmetries viz. $A[P_t^{(2)}]$ and $A[P_W^{(2)}]$ along with the vector-vector correlation $A[PP_{2(1)}^{tW}]$ provides the most dominant sensitivity.  The geometric discord w.r.t both the sub-system, $D_g^t$ and $D_g^W$ offers a comparable sensitivity to that of vector polarization of both top quark and $W$ boson viz. $A[P_2^t]$ and $A[P_2^W]$. The geometric discords offers a higher sensitivity to that of spin correlation parameter $A[PP_{2(1)}^{tW}]$. For both the anomalous moments, the negativity $\mathcal{C}^{\mathrm{PPT}}$ provides a least sensitivity.
\begin{figure*}[!htb]
    \centering
    \includegraphics[width=0.49\textwidth]{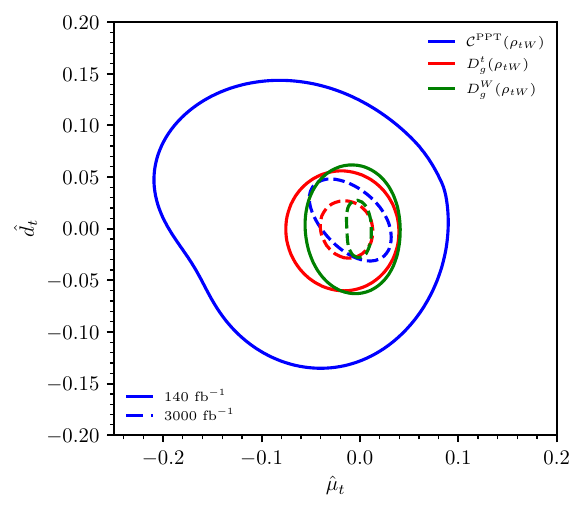}
    \includegraphics[width=0.49\textwidth]{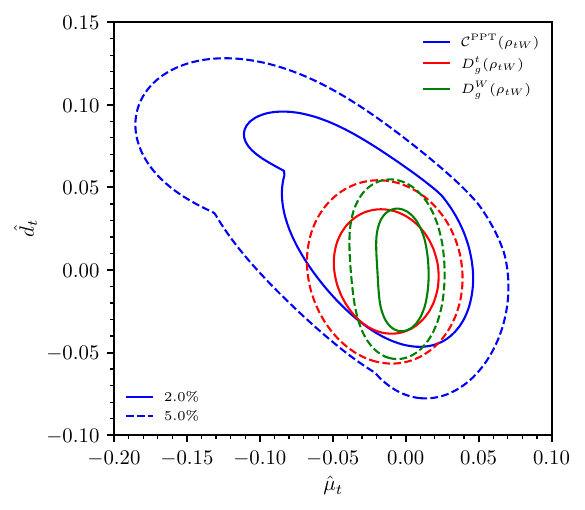}
    \caption{Two-dimensional $95\%$ CL contours for three quantum observables in the anomalous top-quark dipole-moment plane. Left: zero systematic uncertainty; solid and dashed curves correspond to $\mathcal{L}=140$ and $3000~\mathrm{fb}^{-1}$, respectively. Right: $\mathcal{L}=3000~\mathrm{fb}^{-1}$ with benchmark systematic uncertainties of $2\%$ (solid) and $5\%$ (dashed). The likelihood profiles combine the five $m_{tW}$ regions of Table~\ref{tab:sment} at the $\chi^2$ level.}
    \label{fig:2d_lumi_syst}
\end{figure*}

We also compare the reach of sensitivity of QIs to $\hat{\mu}_t$ and $\hat{d}_t$ at projected HL-LHC integrated luminosity of $3000$ fb$^{-1}$ to that of current LHC luminosity. For that, in Fig.~\ref{fig:2d_lumi_syst} (left panel), we present a two dimensional $95\%$ confidence level (CL) contours for all three QIs at $\mathcal{L}=3000$ fb$^{-1}$ and $\mathcal{L}=140$ fb$^{-1}$ as a function of two anomalous couplings. The negativity shows a most enhanced effect on increased luminosity, where the contours shrink by a factor of $\sim 4$, while the contours for geometric discord shows a reduction by a factor of $2.5-4$ on increasing $\mathcal{L}$ from $140$ fb$^{-1}$ to the HL-LHC scenario.

Further, to estimate the impact of systematic uncertainty on the sensitivity of QIs, we compute the $95\%$ CL contours (shown in right panel of Fig.~\ref{fig:2d_lumi_syst}) for two flat benchmark values of systematics i.e, $2\%$ (solid curve) and $5\%$ (dashed curve) at $\mathcal{L}=3000$ fb$^{-1}$. The inclusion of systematic are done as defined in Eq.~\eqref{eq:syst_err}. We observe an increment of contour area by a factor of $\sim 2$ for all three QIs on increasing the systematics from $2\%$ to $5\%$.

We have shown that negativity and geometric discord can be reliably extracted for the $tW$ system. This is a mixed-dimensional $\mathcal{H}_2 \otimes \mathcal{H}_3$ state in a di-leptonic final state, with the neutrino momentum reconstructed using machine-learning techniques. Both observables respond systematically to anomalous top-quark couplings. Notably, the geometric discord $D_g^t$ and $D_g^W$ reaches a sensitivity comparable to that of the vector polarizations $A[P_2^t]$ and $A[P_2^W]$ and exceeds the
spin-correlation parameter $A[PP_{2(1)}^{tW}]$, while the negativity $\mathcal{C}^{\mathrm{PPT}}$ remains the least sensitive for both anomalous moments.

Several avenues for improvement remain open. Next-to-leading-order (NLO) corrections to the quantum-information observables are important: at NLO in QCD, the $t\bar{t}$ and $tW$ amplitudes interfere, and a consistent treatment requires dedicated diagram
removal or diagram subtraction techniques. The increased $t\bar{t}$ cross section at NLO would further dilute $\rho_{tW}^{\prime}$ and reduce the reconstructed correlations, potentially softening the sensitivity reported here. In the current work, we consider a flat systematic uncertainties; a proper treatment of detector effects, parton distribution functions, and jet reconstruction would be required for a realistic projection and lies beyond the scope of this analysis.


\section{Conclusion and Summary}
\label{sec:conclusion}
We have studied non-classical correlations in the mixed-dimensional qubit--qutrit system realized by leading-order $tW$ production at the $13$~TeV LHC in the dileptonic final state. The top-quark spin and $W$-boson polarization define a bipartite state in
$\mathcal{H}_2\otimes\mathcal{H}_3$, whose production density matrix is reconstructed in a generalized Bloch representation from the joint angular distributions of the charged-lepton decay products. The two missing neutrinos, which otherwise obstruct an event-by-event reconstruction of $\rho_{tW}$, are inferred using the conditional normalizing-flow network $\nu^2$-Flows, which reproduces the truth-level neutrino kinematics with sub-GeV bias and correlation coefficients of $0.41$--$0.57$, with the largest residual uncertainty confined to the longitudinal momentum component.

Non-classical correlations in $\rho_{tW}$ are quantified through the Peres--Horodecki negativity $\mathcal{C}^{\mathrm{PPT}}$ and the geometric quantum discord $D_g$ with respect to both subsystems, for which we use the closed-form expression valid for arbitrary bipartite dimension given in Ref.~\cite{Loubenets:2024opb}. Within the Standard
Model, all three observables are non-zero across the full $m_{tW}$ spectrum and grow with the invariant mass, roughly doubling between the lowest and highest mass regions considered, with $D_g^t$ exhibiting the steepest rise. Neutrino reconstruction systematically dilutes $\mathcal{C}^{\mathrm{PPT}}$ at low $m_{tW}$, while the discord
observables are comparatively more robust and even mildly overestimated in the high-mass bins; in all cases the fractional bias shrinks to $\lesssim 20\%$ in the region $m_{tW}\gtrsim600$~GeV, which is also where sensitivity to higher-dimensional operators is largest. Including the dominant reducible $t\bar{t}$ and $WWj$ backgrounds,
tagged with a gradient-boosted decision tree, dilutes the reconstructed observables further: negativity is suppressed by up to $\sim30\%$ in the intermediate-mass bins, while the discords are diluted by $\sim 10$--$25\%$ over most of the spectrum, with all three observables settling to a common $\sim 10$--$15\%$ dilution in the high-mass tail.
This indicates that discord-type correlations are a more resilient probe of non-classicality in the presence of contamination.

We further examined the response of these observables to the anomalous chromomagnetic and chromoelectric dipole moments of the top quark, $\hat{\mu}_t$ and $\hat{d}_t$. The negativity displays a concave profile peaked away from the SM point, while both discords are convex, indicating that the leading quadratic dependence on $\hat\mu_t$ and
$\hat d_t$ has opposite sign for the two classes of observables. A $\Delta\chi^2$ analysis at $\mathcal{L}=140$~fb$^{-1}$ shows that $D_g^W$ provides the dominant sensitivity to $\hat{\mu}_t$ among the three quantum observables, followed by $D_g^t$, while for $\hat{d}_t$ the two discords give comparable constraints. The negativity
$\mathcal{C}^{\mathrm{PPT}}$ remains the least sensitive for both couplings. The high-$m_{tW}$ bin M5 dominates the combined constraint in nearly all cases. A comparison with the constituents of the density matrix shows that the polarization asymmetries $A[P_t^{(2)}]$ and $A[P_W^{(2)}]$, together with the vector--vector correlation
$A[PP_{2(1)}^{tW}]$, provide the strongest overall sensitivity. The
geometric discord of both subsystems reaches a sensitivity comparable
to that of the vector polarizations and exceeds that of the spin-correlation parameter,
while the negativity remains the least sensitive for both anomalous moments. At the projected HL-LHC luminosity of $3000$~fb$^{-1}$, the two-dimensional exclusion contours shrink by a factor of $\sim4$ for the negativity and by $2.5$--$4$ for the discords relative to current LHC statistics. Increasing the systematic uncertainty from $2\%$ to
$5\%$ enlarges the $95\%$ CL contours in the $(\hat{\mu}_t,\hat{d}_t)$ plane by approximately a factor of two.

The central result of this work is that genuinely mixed-dimensional, qubit--qutrit quantum correlations in the $tW$ system can be reconstructed from realistic LHC final states, survive both neutrino-reconstruction and dominant-background effects, and respond
systematically to anomalous top-quark couplings. In particular, the geometric discord provides a sensitivity competitive with the vector polarization asymmetries of the top quark and $W$ boson, establishing quantum-information observables as viable probes of new physics in mixed-dimensional systems. A full treatment including NLO
$t\bar{t}$--$tW$ interference and detector-level systematics is left for future work.

\section*{Acknowledgements}

This work was supported by  National Natural Science Foundation of China under Grant Nos. T2241005 and 12075059 and by startup fund of USTC.

\bibliographystyle{JHEP} 
\bibliography{refer.bib}

\end{document}